\documentclass[aps,nofootinbib,superscriptaddress,twocolumn,prl,longbibliography]{revtex4-1}
\usepackage{amssymb}
\usepackage{natbib}
\usepackage{amsmath}
\usepackage{amsfonts}
\usepackage{graphicx}
\usepackage{mathrsfs}
\usepackage{dcolumn}
\usepackage{bm}
\usepackage{color}
\usepackage{cancel}
\usepackage{mathbbol}
\usepackage{epsfig}
\usepackage{units}
\usepackage{esint}
\usepackage{soul}
\usepackage{color}

\definecolor{rot}{rgb}{0.75,0.05,0.25}
\definecolor{hellgrau}{gray}{0.5}
\definecolor{blau}{rgb}{0,0,0.7}

\def\Tr{\mbox{Tr}}

\begin{document}

\title{The power of a critical heat engine}

\author{Michele Campisi}
\affiliation{NEST, Scuola Normale Superiore \& Istituto Nanoscienze-CNR, I-56126 Pisa, Italy}
\email{michele.campisi@sns.it}
\author{Rosario Fazio}
\affiliation{NEST, Scuola Normale Superiore \& Istituto Nanoscienze-CNR, I-56126 Pisa, Italy}
\affiliation{ICTP, Strada Costiera 11, 34151 Trieste, Italy}

\date{\today }

\begin{abstract}
Since its inception about two centuries ago thermodynamics has sparkled continuous interest and fundamental questions. According to the second law no heat engine can have an efficiency larger than Carnot's efficiency. The latter can be achieved by the Carnot engine, which however ideally operates in infinite time, hence delivers null power. A currently open question is whether the Carnot efficiency can be achieved at finite power. Most of the previous works addressed this question within the Onsager matrix formalism of linear response theory. Here we pursue a different route based on finite-size-scaling theory. We focus on quantum Otto engines and show that when the working substance is at the verge of a second order phase transition diverging energy fluctuations can enable approaching the Carnot point without sacrificing power. The rate of such approach is dictated by the critical indices, thus showing the universal character of our analysis. 
\end{abstract}

\maketitle

\section{Introduction}
\begin{figure}[b]
\includegraphics[width=\linewidth]{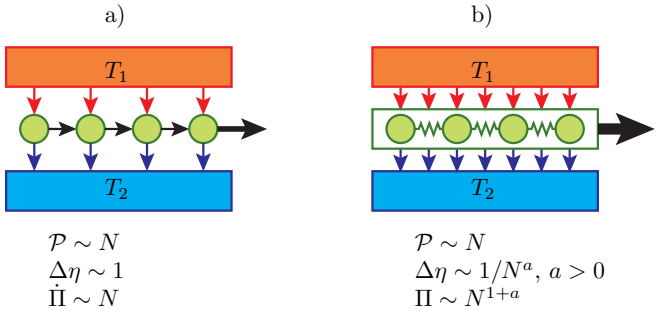}
\caption{Main idea at the basis of the results. Panel a) $N$ identical devices operating in parallel provide a power $\mathcal P$ that scales like $N$, at fixed efficiency.  Panel b) When an interaction among the $N$ parallel devices is turned on the approach to Carnot efficiency is enabled.}
\label{fig:1}
\end{figure}
Increasing the power output of a heat engine has a corresponding cost in terms of reduced efficiency $\eta$; or, in equivalent terms, 
a larger deviation $\Delta \eta= \eta^{\rm C}- \eta$ from the Carnot efficiency $\eta^{\rm C}$.  A question that is currently the object of vigorous research efforts is whether it is possible to devise a heat engine that outputs  finite power at Carnot efficiency \cite{Benenti11PRL106,Allahverdyan13PRL111,Proesmans15PRL115,Brandner15PRX5,Polettini15PRL114}. In the following we address this question by focussing on the the following quantity
\begin{align}
\dot \Pi \doteq \frac{\mathcal P}{\Delta \eta} 
\end{align}
that is  the ratio of power output $\mathcal P$ over $\Delta \eta$. We shall call $\dot \Pi$ the performance rate.
It is trivially possible to increase the power without affecting the efficiency by scaling the size of the working substance: An array of $N$ 
identical engines working in parallel provides  an $N$-fold larger power than each of them, at the same efficiency. 
In this case the output work per cycle and efficiency scale as  $\mathcal P \sim N$,  and $\Delta \eta \sim 1$, consequently
$\dot  \Pi  \sim N $ (here the symbol $\sim$ means ``scales as''). Note that this linear increase in performance rate does not represent any real gain as it is achieved at the cost of a corresponding  linear increase of resources. The question is therefore ``Can the scaling of the performance rate $\dot \Pi$ be improved beyond linear'' in order to have a true gain? Note that a positive answer implies (in an asymptotic sense) a positive answer to the fundamental question posed above: Imagine to have a working substance made up of $N$ constituents (or resources); if $\dot \Pi \sim N^{1+a}$, with $a>0$, then one could  approach Carnot efficiency as $\Delta \eta \sim \mathcal P / \dot \Pi \sim N^{-a} \rightarrow 0$ by increasing $N$, while keeping the ``power per resource'' fixed i.e., $\mathcal P \sim N$.

\begin{figure}
\includegraphics[width=\linewidth]{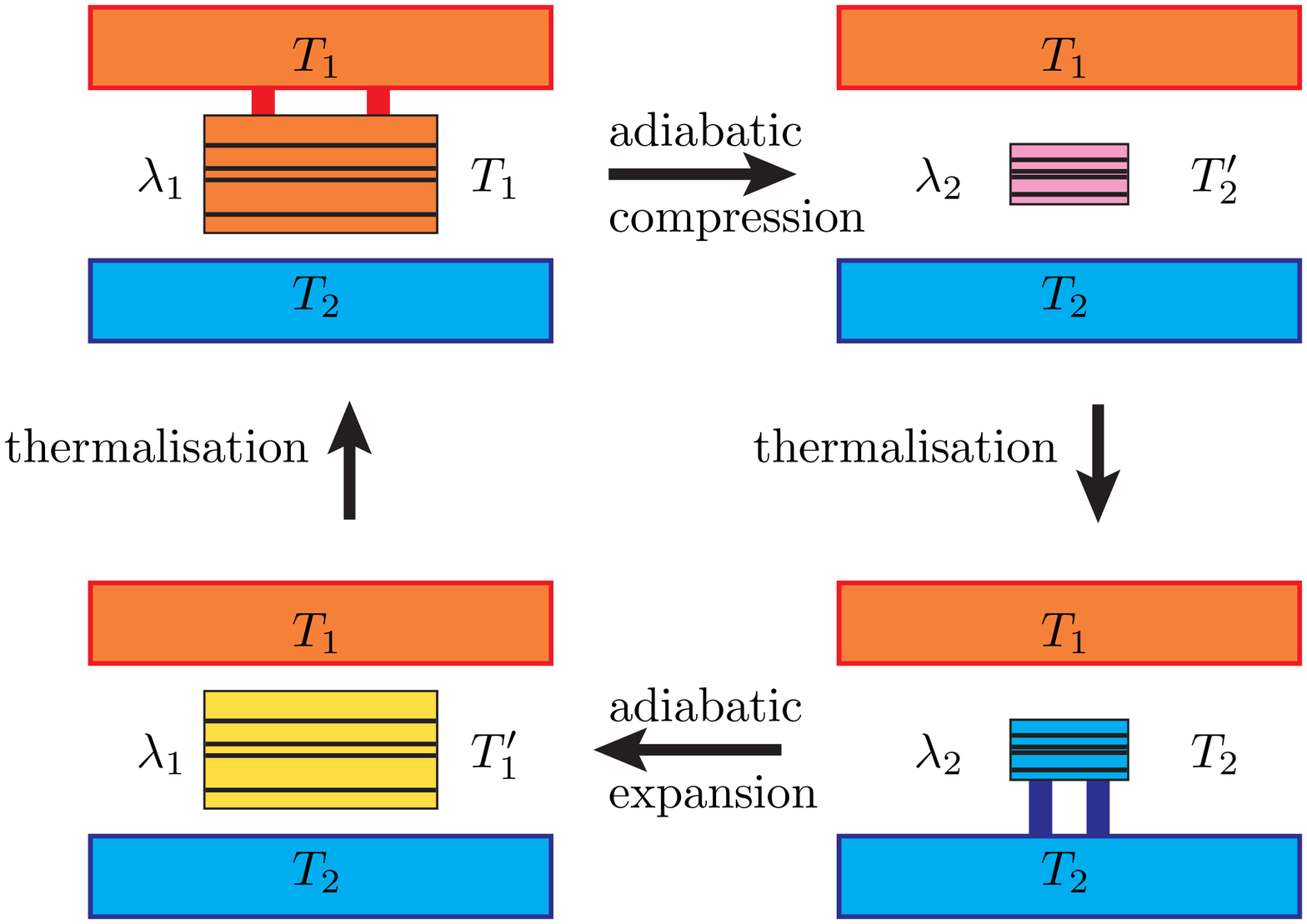}
\caption{A quantum Otto engine. At time $t=0$ the WS is in thermal equilibrium with bath $1$ and $\lambda=\lambda_1$. During the first stroke the WS undergoes a thermally 
isolated transformation where the Hamiltonian switches from $H_1^\text{WS}=\lambda_1 K$ to $H_2^\text{WS}=\lambda_2 K$. The second 
stroke consists in thermalisation with bath $2$, at $\lambda$ being kept fixed at $\lambda_2$. During the third  stroke the Hamiltonian goes 
back to $H_1^\text{WS}=\lambda_1 K$ while in thermal isolation. The fourth stroke consists in letting the system thermalise with bath $1$, 
with $\lambda$ being kept fixed at $\lambda_1$.}
\label{fig:2}
\end{figure}

In the following we make a substantial step towards answering the question above. The main idea that we pursue is that an interaction between the $N$ constituents of the working substance could provide the extra scaling power that is needed for a positive resolution, see Fig. \ref{fig:1}.
 In order to address that question quantitatively we consider a special engine 
cycle that is well studied in the literature (see, e.g. \cite{Benenti13arXiv13114430,Kosloff14ARPC65,GelbwaserKlimovsky15AAMOP64} and reference therein), namely a quantum version of the Otto engine~\cite{Scully02PRL88,Feldmann03PRE68,Quan07PRE76,Niskanen07PRB76,Allahverdyan08PRE77,Rossnagel14PRL112},
see Fig. \ref{fig:2}. We show that 
a universal behaviour, with anomalous scaling of the performance rate 
\begin{align}
\dot \Pi  \sim N^{1+(\alpha-z\nu)/(d\nu)}
\label{eq:PiDot}
\end{align}
emerges  when the working substance is on the verge of a second order phase transition.
Here $\alpha, \nu, z$ are the specific heat, correlation length and dynamical critical exponent, and 
$d$ is the number of dimensions of the working substance.
Note that the performance rate contains two concurrent contributions, one stemming from the scaling of the
heat capacity, i.e. the exponent $\alpha$, and one stemming from the scaling of the
relaxation time, i.e., the exponent $z$.  Since $d$,$\nu>0$, in order to have a more than linear behaviour of $\dot \Pi$ one needs a  working 
substance with a critical point characterised by the inequality $\alpha-z \nu>0$. In fact as we explain below
in detail, the stronger condition 
\begin{align}
\alpha-z \nu\geq1
\label{eq:condition}
\end{align}
would ensure the asymptotic approach toward the Carnot point without giving up power per resource.
Is that possible?  The finiteness of internal energy implies the bound $\alpha \leq 1$. Critical slowing down ($z  \geq 0$) would then imply $\alpha-z \nu \leq 1$ (e.g. in the three-dimensional Ising model, $\alpha \simeq 0.12, \nu \simeq 0.63$~\cite{Pelissetto02PHYSREP368}, and $z \simeq 2.35$ \cite{Matz94JSP74} hence $\alpha-z \nu\simeq {-0.28}$). A number of theoretical and experimental works report on the possibility of the exotic phenomenon of critical speeding up $z\leq0$ \cite{Zappoli90PRA41,Boukari90PRL65,Grams14NATCOMM5,Tavora14PRL113}. In this case the two terms might add up above unity (e.g., recent experimental studies report $\alpha \simeq 0.38$ \cite{Higashinaka04JPSJ73}  and $z\nu \simeq-0.7$ \cite{Grams14NATCOMM5}, hence $\alpha-z \nu \simeq 1.08 >1 $, for $\rm{Dy_2Ti_2O_7}$). We conclude that there currently appear to be no fundamental reason hindering the possibility of approaching Carnot efficiency without giving up power per resource. This is certainly possible up to some threshold size $\bar N$ for which the weaker condition $\alpha-z \nu>0$ suffices.
Below we illustrate how such powerful critical engines should be designed.  At the heart of our result is the recognition that scaling of the performance, i.e. 
\begin{align} 
\Pi = \frac{W_\text{out}}{\Delta \eta}
\end{align} 
where $W_\text{out}$ is the work output per cycle, is dictated by the heat capacity of the working substance (which can notably diverge at the critical point), a crucial and simple fact that was never noticed before, see Eq. (\ref{eq:dWdlambda}) below.

\section{Results}
\subsection{N-body quantum Otto engine}
The quantum Otto engine (see Fig.\ref{fig:2}) is a four-stroke engine based on a working-substance (WS) with Hamiltonian
\begin{align}
H^\text{WS}(t)=\lambda(t) K.
\label{eq:Hi}
\end{align} 
(Following the 
current literature we call these engines ``quantum Otto engines'' although they bear no other quantum feature  besides the discreteness of the spectrum. Accordingly there is nothing genuinely quantum in our treatment).
The heats released in the baths during the thermalisation strokes are 
$Q_i= \mp \lambda_i [U_K(\beta_2\lambda_2)-U_K(\beta_1\lambda_1)]$, where 
$
U_K(\theta) = \Tr \, K \, e^{-\theta K} / \Tr \, e^{-\theta K}
$
is the internal energy associated to the base Hamiltonian $K$ at inverse temperature $\theta$, and $-,+$ is for $i=1,2$ respectively. 
The work performed by the engine $W_\text{out}= Q_1+Q_2$ is thus 
\begin{align}
W_\text{out}= (\lambda_2-\lambda_1) [U_K(\beta_2\lambda_2)-U_K(\beta_1\lambda_1)]. \label{eq:W}
\end{align}
For $\beta_1<\beta_2$, the condition $W_\text{out} \ge 0$, $Q_1 \ge 0$ defines the regime of operation of the engine as a heat engine, implying 
$\lambda_2/\lambda_1 \geq \beta_1/\beta_2$. The  efficiency is 
\begin{align}
\eta = \frac{W_\text{out}}{Q_1} = 1-\frac{\lambda_2}{\lambda_1} \leq 1-\frac{\beta_1}{\beta_2}= \eta^{\rm C}
\label{eq:eta}
\end{align}
The efficiency is smaller than $\eta^{\rm C}$ in accordance with the heat engine fluctuation relation~\cite{Campisi14JPA47,Campisi15NJP17}. 

\begin{figure}
\includegraphics[width=\linewidth]{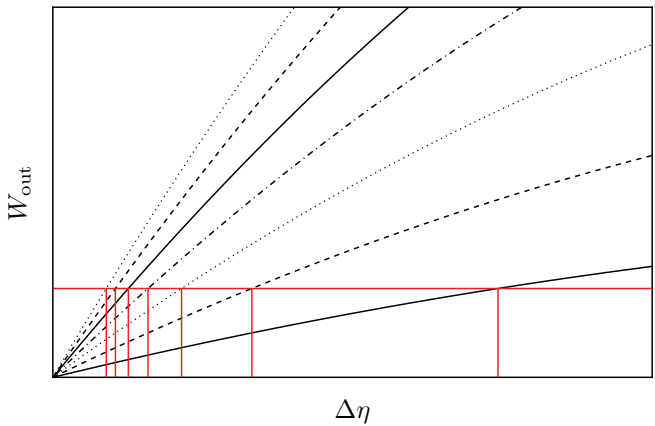}
\caption{Graphical demonstration of Equation (\ref{eq:dWdlambda}). The graph shows various plots of $W_\text{out}(\Delta \eta)$ for increasing values of $N$ (from lower to upper curves). At a fixed work output (horizontal red line), the curves are intercepted at decreasing values of $\Delta \eta$.  If the slope at the origin increases with a power $N^a$, $\partial W_\text{out}/{\partial \Delta \eta} |_{\Delta \eta=0} \sim N^a$, then, provided for larger and larger $N$ the intercept occurs in the region where the linear approximation holds for the $N^\text{th}$ curve, the approach towards $\Delta \eta = 0$ occurs as $1/N^{a}$, hence $\Pi=W_\text{out}/\Delta \eta \sim N^a \sim \partial W_\text{out}/{\partial \Delta \eta} |_{\Delta \eta=0}$. Similarly, one might rise (lower) the value of $W_\text{out}$ with $N$, e.g., as $W_\text{out}(\Delta \eta)= w N^b$. Accordingly, provided for larger and larger $N$ the intercept still occurs in region where the linear approximation holds, the approach towards $\Delta \eta = 0$ occurs as $1/N^{a-b}$. Still, $\Pi=W_\text{out}/\Delta \eta \sim N^a \sim \partial W_\text{out}/{\partial \Delta \eta} |_{\Delta \eta=0}$. 
}
\label{fig:3}
\end{figure}

With Eq. (\ref{eq:eta}) the output work  of the quantum Otto engine can be expressed as a function of $\lambda_1,\beta_1,\beta_2$ and $\Delta \eta=\lambda_2/\lambda_1-\beta_1/\beta_2$ as
\begin{align}
W_\text{out}= \lambda_1(\Delta \eta -\eta^{\rm C}) [U_K(\beta_1\lambda_1+\beta_2 \lambda_1\Delta \eta)-U_K(\beta_1\lambda_1)]\label{eq:W-DeltaEta}
\end{align}
This expression allows us to study the performance $\Pi$.
In the region where linear approximation hold, i.e., $\Delta \eta \ll 1$ (namely $\beta_1\lambda_1$ is close to $ \beta_2 \lambda_2$),
the scaling of the engine's performance $\Pi$ with $N$ is given by the scaling of 
${\partial W_\text{out}}/{\partial \Delta \eta}  | _{\Delta \eta=0}$, i.e. the slope of 
$W_\text{out}(\Delta \eta)$ at the origin. This is illustrated in Fig.\ref{fig:3}.
In the case of $N$ devices in parallel it is $W_\text{out}= N W_\text{out}^\text{single}$ (here $W_\text{out}^\text{single}$ denotes 
 the work of each single device), hence trivially ${\partial W_\text{out}}/{\partial \Delta \eta}  | _{\Delta \eta=0} \sim \Pi \sim N$,  and, accordingly $\dot \Pi \sim N$, as we noted before. It follows that in order to boost the scaling of the performance it is 
 necessary to have the slope  ${\partial W_\text{out}}/{\partial \Delta \eta}  | _{\Delta \eta=0}$ to scale more than linearly.

By taking the derivative of Eq. (\ref{eq:W-DeltaEta}) with respect to $\Delta \eta$ we obtain the central relation 
 \begin{align}
\Pi \sim \frac{\partial W_\text{out}}{\partial \Delta \eta} |_{\Delta \eta=0} \sim N c_K(\beta_1 \lambda_1) 
\label{eq:dWdlambda}
\end{align}
where  $c_K(\theta)= -(1/N)\theta^2\partial U_K/\partial \theta$ is the specific heat of the working substance. 
To understand the physics behind the emergence of Eq. (\ref{eq:dWdlambda}), consider working at some point that is very close to the Carnot point, i.e. chose $\lambda_1$ and $\lambda_2$ so that their ratio $\lambda_1/\lambda_2$ is very close to $\beta_2/\beta_1$. After the adiabatic compression stroke the working substance reaches a new temperature $T'_2= 1/(k_B \beta'_2)$ that is very close to the cold bath temperature $ T_2 =1/(k_B \beta_2)$: $T'_2 = T_1 \lambda_2/\lambda_1 = T_2  + \Delta T_2$, with a small $\Delta T_2$. The larger the heat capacity $C_K=N c_K$ of the working substance, at the thermalisation point, the larger the heat exchanged $Q_2 = C_K \Delta T_2$ during the subsequent thermalisation with the cold bath. Likewise for the subsequent expansion and thermalisation. Since $W_\text{out} = Q_1+Q_2$, the larger the heat capacity the larger the work output, and accordingly the larger the performance.

Eq. (\ref{eq:dWdlambda}) tells us that in order to achieve super-linear scaling of the performance, one needs a working substance with an anomalous scaling of the specific heat. Recall that for ordinary substances, the heat capacity is extensive $C_K\sim N$, namely $c_K=C_K/N\sim 1$. What is needed for improved performance is $c_K\sim N^a$, with some $a>0$. This can happen at the verge of a phase transition. The physical reason is that at a phase transition finite exchanges of heat are accompanied by infinitesimal changes of temperature, i.e., the specific heat diverges. This is because at the transition point the energy intake is not employed to heat up but rather to make the change of phase.

For a second order phase transition, finite size scaling predicts a peak in the specific heat whose 
height and width scale respectively as $\bar{c}_K\sim N^{\alpha/(d\nu)}$ and $\delta \sim N^{-1/(d\nu)}$ \cite{Fisher72PRL28}, 
where $d$ is the dimensionality of the system, $\alpha$ and $\nu$ are the specific heat and the correlation length 
critical exponents. Accordingly we predict the possibility of a boosted scaling of the perfmormance $\Pi \sim N^{1+\alpha/(d\nu)}$.
Writing the performance rate as $\dot \Pi = \mathcal P / \Delta \eta= W_\text{out}/( \mathcal T \Delta \eta)=\Pi/\mathcal T$ we see that its scaling is determined by the scaling of $\Pi$ and of  the cycle time $\mathcal T$. The latter
is dominated by the thermalisation time \cite{Allahverdyan13PRL111}, which, according to finite size scaling theory scales with the dynamical critical exponent $z$ as $\mathcal T_\text{relax} \sim N^{z/d}$~\cite{Suzuki77PTP58}. This gives equation \ref{eq:PiDot}.

\subsection{Critical engine design}
We illustrate how a quantum Otto engine can be designed to achieve the predicted performance rate in Eq. \ref{eq:PiDot}. Let the substance be described by a Hamiltonian $K$, displaying, in the infinite size limit, a second order phase transition at the inverse critical temperature $\theta_{\rm C}$. Let the two baths have the inverse temperatures $\beta_1,\beta_2$. We assume we can stretch/compress the spectrum of the working substance and implement accordingly the time dependent Hamiltonian in Eq. (\ref{eq:Hi}) with $\lambda(t)\in[\lambda_1,\lambda_2]$. When $\lambda$ takes on the value $\lambda_i$, the critical temperature is  rescaled to $\beta_{\rm C}^i=\theta_{\rm C}/\lambda_i$. We choose, e.g., $\lambda_1$ so that $\beta_1 = \beta_{\rm C}^1$ (that is $\lambda_1=\theta_{\rm C}/\beta_{\rm C}^1$). In this way the temperature of the bath $1$ coincides with the critical temperature of $H_1^\text{WS}=\lambda_1K$. We next choose $\lambda_2$ as the solution of $W_\text{out}=N^{1+z/d}\, w $, for some fixed $w$. That gives the corresponding efficiency $\eta = 1-\lambda_2/\lambda_1$. Note that since $\mathcal T \sim N^{z/d}$, the power per constituent is fixed: $\mathcal P = W_\text{out}/\mathcal T \sim N$.
Since the slope of the graph $W(\Delta \eta)$ grows with the heat capacity as $\Pi \sim N^{1+\alpha/(d\nu)}$, we have $\Delta \eta =W_\text{out}/\Pi \sim N^{1+z/d}/ N^{1+\alpha/(d\nu)} \sim N^{-(\alpha-\nu z)/d \nu} $. Hence
if $\alpha-\nu z > 0$ the solution $\lambda_2$ gets  closer and closer to $\lambda_1\beta_1/\beta_2$. Accordingly the efficiency $\eta=1-\lambda_2/\lambda_1$ gets closer and closer to $\eta^{\rm C}=1-\beta_1/\beta_2$. As discussed in the caption of Fig. \ref{fig:3} the previous statement is correct as long as the $\lambda_2$ falls in the region of validity of the linear approximation (that is, in graphical terms, if the corresponding $\Delta \eta$ falls in the region where the curve $W(\Delta \eta)$ is well approximated by a straight line passing through the origin). The extension of that region corresponds to the width of the specific heat peak, which as mentioned above, universally scales as $\delta \sim N^{-1/(d\nu)}$. So, in order to keep pace with the shrinking of the linear approximation region, the approach towards $\Delta \eta =0$ should be not slower than $N^{-1/(d\nu)}$. 
Since that would occur at a pace of $N^{-(\alpha-\nu z)/(d\nu)}$, the condition ensuring that it actually occurs is equation (\ref{eq:condition}).
\begin{figure}
\includegraphics[width=\linewidth]{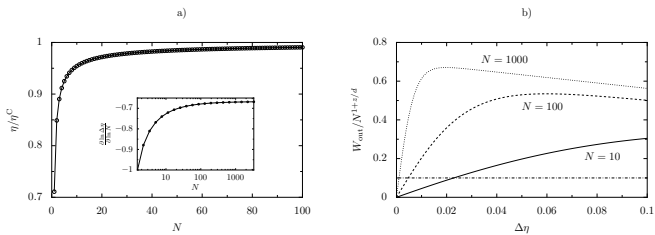}
\caption{Efficiency and shrinking of linear region in a critical sped-up engine.  Panel a). Approach towards the Carnot point of a critical working substance characterised by Eq. (\ref{eq:condition}) at fixed power per constituent. Plots were calculated with critical indices reported recently for $\rm{Dy_2Ti_2O_7}$, namely $\alpha \simeq 0.38$ \cite{Higashinaka04JPSJ73}  and $z\nu \simeq-0.7$ \cite{Grams14NATCOMM5}. The value of $\nu$ was obtained from the relation $\nu= (2-\alpha)/d$ \cite{Huang87Book}, Carnot point was at $\eta^{\rm C}=1/2$ and $w=W_\text{out}/N^{1+z/d}=0.1$. The inset shows the quantity $\partial \ln \Delta \eta/\partial \ln N$ as a function of $N$, that is the exponent characterising the approach to the Carnot point that reaches the expected value $-(\alpha - z \nu)/(d \nu)\simeq -0.67$. Panel b) Rescaled output work $W_\text{out}/N^{1+z/d}$ as a function of $\Delta \eta$. Note that the linear region shrinks around the origin of axes as $N$ increases, and that the intercept with $W_\text{out}/N^{1+z/d}=0.1$ occurs within that region for all curves.
}
\label{fig:4}
\end{figure}

In Fig. \ref{fig:4}.a we illustrate how in such a case, the engine design that we have described above actually results in an asymptotic approach towards the Carnot point, at fixed power per constituent. Fig. \ref{fig:4}.b illustrates the shrinking of the linear region, due to the narrowing of the peak width.
In making Fig. \ref{fig:4} we have taken full advantage of one of the most striking aspects of our analysis, namely its universality. The details of the specific model are not essential, all that counts are the critical exponents. The specific heat peak reads $
c_K(\theta) =N^{ \alpha /(d \nu)} \theta^2 U_1 \Delta U f \left (\Delta U (\theta - \theta_{\rm C}) N^{1/(d \nu)}\right)
$ where $f(x)$ is some bell shaped function (its exact shape is not relevant). The according internal energy of the working substance around the critical point reads
$
U_K(\theta) = U_0 - N^{1 + (\alpha - 1)/(d \nu)} U_1 F \left(\Delta U(\theta-\theta_{\rm C}) N^{1/(d \nu)}\right) \, .
$
where $F'(x)=f(x)$. The plots are obtained by using  the latter with Eq. (\ref{eq:W}) to evaluate the work output, with the choice $f(x)=\text{sech}^2(x), F(x)=\tanh(x)$ and $U_1=\Delta U_1=1$. The plot illustrates the approach towards Carnot efficiency at fixed power per constituent, with the predicted scaling exponent $-(\alpha - z\nu)/(d\nu)$.

\section{Discussion}
The idea that phase transitions could enable the attainment of Carnot efficiency at finite power was previously hinted by Polettini \emph{et al.} \cite{Polettini15PRL114}, but was never pursued before. The present work confirms that intuition in fully fledged way based on universality and finite-size scaling theory and most importantly by accounting for the first time for the effect of criticality on the time of operation hence of the power.

As is clear from Fig. \ref{fig:3}, independent on how the slope  ${\partial W_\text{out}}/{\partial \Delta \eta}  | _{\Delta \eta=0}$ scales, it is $W_\text{out}(\Delta \eta =0)=0$, hence exactly at the Carnot point all quantum Otto engines deliver null work hence null power. Our statement should be accordingly understood in a weaker asymptotic sense, namely, that it is possible to get as close as one wants to the Carnot point without giving up power per constituent. This is the viewpoint that also inspires Ref. \cite{Allahverdyan13PRL111}.

We remark that our linear condition $\Delta \eta \ll1$ substantially differs from the condition of linear response regime (i.e. $\beta_2 -\beta_1\ll\beta_2$ or $\eta \ll 1$) that was investigated previously in \cite{Benenti11PRL106,Proesmans15PRL115,Brandner15PRX5,Polettini15PRL114}. In our case the two temperatures need not be close to each other.

We stress that all obstacles hindering the realisation of the critical powerful Carnot engines appear to be of technological nature, rather than fundamental. One major difficulty stems from the necessity of implementing the Hamiltonian (\ref{eq:Hi}) containing a global coupling $\lambda(t)$, which can be very challenging in practice.
Another difficulty stems from the fact that in order to be able to asymptotically approach the Carnot point, one should accordingly have an increasing degree of accuracy with which $\lambda_1$ and $\lambda_2$ are controlled. 

Lastly it is worth stressing that the result in Eq.  (\ref{eq:dWdlambda}) holds in general as a consequence of the linear approximation valid for $\Delta \eta \ll 1$, and is accordingly not restricted to the case of critical phenomena. This tells that one route toward the improvement of the performance of a working substance is to increase its specific heat. That could be achieved by increasing its number of constituents, a possibility that we have investigated here, or by manipulating any other parameter entering the Hamiltonian $K$. Recently reported cases  of improved performances  \cite{Cakmak15arXiv} can in fact be interpreted in terms of increased heat capacity.

\subsection{Data availability statements}
The authors declare that all data supporting the findings of this study are available within the article.

\subsection{Acknowledgments}
This research was supported by the 7th European Community Framework Programme under grant agreements n. 623085 (MC-IEF-NeQuFlux), 
n. 600645 (IP-SIQS), n. 618074 (STREP-TERMIQ) and by the COST action MP1209 ``Thermodynamics in the quantum regime''.

\subsection{Authors Contribution}
M. C. conceived the idea. M. C. and R. F. carried the work, analysed the results and drew the conclusions.

\subsection{Competing financial interests}

The authors declare no competing financial interests.

\end{document}